\documentclass[aps,onecolumn,superscriptaddress]{revtex4}
\input epsf
\usepackage{epsfig}
\usepackage{amsmath,amssymb,latexsym}
\newcommand{\switchfonts}{\usefont{OT1}{cmr}{m}{it}}
\newcommand{\lie}{\switchfonts \mbox{\symbol{36}} \normalfont}  
\newcommand{\bra}[1]{\langle #1|}
\newcommand{\ket}[1]{|#1\rangle}

\newcommand{\sqrthb}{\sqrt{| \bar{h}| } }

\newcommand{\intsigma}{\int_{\Sigma_{t}}}

\newcommand{\piphi}{\pi_{\phi}}

\newcommand{\sigmat}{\Sigma_{t}}

\newcommand{\nablada}{ \bar{\nabla}_{a} }
\newcommand{\nablaua}{ \bar{\nabla}^{a} }
\newcommand{\nabladb}{ \bar{\nabla}_{b} }

\newcommand{\nabladc}{ \bar{\nabla}_{c} }
\newcommand{\nablauc}{ \bar{\nabla}^{c} }

\newcommand{\nabladm}{ \bar{\nabla}_{m} }
\newcommand{\nablaum}{ \bar{\nabla}^{m} }

\newcommand{\nabladell}{ \bar{\nabla}_{\ell} }
\newcommand{\nablauell}{ \bar{\nabla}^{\ell} }
\newcommand{\nabladi}{ \bar{\nabla}_{i} }
\newcommand{\nablaui}{ \bar{\nabla}^{i} }

\newcommand{\nabladO}{ \bar{\nabla}_{0} }

\newcommand{\bbbox}{ \stackrel{ \bar{} }{\Box} }

\newcommand{\xp}{ x^{\prime} }

\newcommand{\gb}{\bar{g}}

\newcommand{\sqrtgb}{\sqrt{ -|\gb| } }

\begin{document}

\title{ On leading order gravitational backreactions in de Sitter spacetime  }
\author{ B.~Losic }
\affiliation{ Department of Physics \& Astronomy, 
University of British Columbia,
6224 Agricultural Road
Vancouver, B.C. V6T 1Z1
Canada }
\affiliation{ Department of Physics,
P-412, Avadh Bhatia Physics Laboratory
University of Alberta,
Edmonton, Alberta T6G 2J1
Canada}
\author{ W.G.~Unruh }
\affiliation{ Department of Physics \& Astronomy, 
University of British Columbia,
6224 Agricultural Road
Vancouver, B.C. V6T 1Z1
Canada }
\affiliation{ Canadian Institute for Advanced Research, Cosmology and Gravitation Program  }
\email{ blosic@phys.ualberta.ca   ;  unruh@physics.ubc.ca}

\date{February 20, 2006 }

\begin{abstract}

Backreactions are considered in a de Sitter spacetime whose cosmological constant is generated by the potential of scalar field. The leading 
order gravitational effect of nonlinear matter fluctuations is analyzed and it is found that the initial value problem for the perturbed 
Einstein equations possesses linearization instabilities. We show that these linearization instabilities can be avoided by 
assuming strict de Sitter invariance of the quantum states of the linearized fluctuations. We furthermore show that quantum anomalies do not block the invariance requirement.
This invariance constraint applies to the entire spectrum of states, from the vacuum to the excited states (should they exist), and is in that sense much stronger than the 
usual Poincare invariance requirement of the Minkowski vacuum alone.  Thus to leading order in their effect on the gravitational field, the quantum states of the matter and 
metric fluctuations must be de Sitter invariant. 

\end{abstract}

\maketitle
\section{Introduction}

The relative importance of nonlinear quantum fluctuations in the very early universe is still not well-understood. Within the inflationary scenario we have recently suggested 
\cite{Losic:2005vg} that their effect on the gravitational field may in fact dominate that of the linear fluctuations for a wide range of slow-roll parameters 
(for long-wavelengths). However, these and similar 
suggestions are made within the context of calculations which are highly coordinate-dependent and are thus confined to making statements within a very specific coordinate 
choice, with all the problems and advantages that entails. Even if one attempts to eliminate this ambiguity by 
using the so-called gauge-invariant approach, and essentially fixes coordinates via a particular coordinate choice which admits no nontrivial residual freedoms, it is 
difficult to characterize the invariant, relative, importance of nonlinear fluctuations. 

Part of this problem in the context of inflationary scenarios is that the linear perturbative matter sector about an inflating background is non-trivial. However, one can 
consider the well-known limit of a vacuum de Sitter spacetime whose cosmological constant $\Lambda$ is sourced from the {\it constant} potential of a linear scalar field 
$\phi$. Given such a background, vacuum, spacetime the leading order of the matter sector occurs quite naturally at {\it second} order in perturbation theory, owing 
essentially to the quadratic character of the Klein-Gordon stress-energy $T_{ab}$. 
One can then ask what the {\it leading} order gravitational 
response to these {\it nonlinear} matter fluctuations will be, without having to disentangle the effects of the linear matter sector from that of the nonlinear one. 
Furthermore, since the scalar field in the background de Sitter spacetime is a constant, the leading order, nonlinear, fluctuations in the matter sector 
will have a simplified coordinate dependence to second order since the Lie derivative of the background field is zero.

It is well known that the theory of a linear quantum field in flat spacetime is well-formulated, and apart from some critical differences 
stemming from the absence of a preferred vacuum state/ global inertial coordinates, that the same theory on a globally hyperbolic curved spacetime\cite{Wald:1994} is 
well-defined as well. 
However, nonlinear interacting fields even in Minkowski spacetime generally have to be regulated in some way and indeed the 
renormalization ambiguities which arise in such a procedure are well specified as renormalized coupling {\it constants} which appear order by order in perturbation theory. 
Until recently, a much larger renormalization ambiguity appeared inherent in the curved spacetime case because, instead of coupling constants, there appeared coupling 
{\it functions}, whose dependence on the spacetime point is wholly arbitrary.  In a series of recent advances, Hollands, Wald and others showed 
(see e.g.  \cite{Hollands:2001qe}, \cite{Hollands:2001fb}, \cite{Hollands:2004yh}, and \cite{Hollands:2002ux}) that the imposition of certain key requirements 
(namely, `locality` and `covariance`) can reduce the renormalization ambiguity in curved spacetime to that of Minkowski spacetime together with that of some additional 
parameters associated with the couplings of the quantum field to curvature.  Although we refer to the reader to the references for the technical details,
the requirement of `locality` can be usefully thought of as a restriction on the regularized quantum fields to be independent of globally defined structures 
(such as a preferred vacuum state) and non-covariant structures like a preferred coordinate system. This use of 'locality' is to be distinguished from the `locality` which 
is typically defined to imply that fields commute for spacelike separations. 

While the linear matter (scalar field) theory on the background de Sitter spacetime is well-defined and has a rich set of possible states, we wish to begin to examine 
the effects that the nonlinearities and in particular the coupling of the scalar field to gravity, has on the theory. The result is surprising. The very requirement that 
one be able to couple the scalar field to gravity in and of itself places severe constraints on the allowed states for the scalar field.  This paper will use the 
'linearization stability constraints' \cite{D'Eath:1976}, \cite{Arms:1977}, \cite{F&M}, \cite{BrillDeser:1973}
(i.e. constraints on the first order perturbation which follow from demanding that the second order 
perturbations have a solution) to argue, in the spirit of previous work by Moncrief \cite{Moncrief:1978te} and Higuchi \cite{Higuchi:1991tk}, \cite{Higuchi:1991tm}, that
all the quantum states (not just the vacuum) of the linear, scalar, quantum field must be de Sitter invariant. We also show that this conclusion is robust under regularization 
- that no anomalies, either in the coordinate conditions, conservation of stress-energy, or `linearization stability constraints` destroy this conclusion. We emphasize that the 
gauge fixing procedure for the gravitational sector we outline in this paper is not necessary for the linearization stability arguments we make, 
the main result of this paper, to go through. However, we do show that one can make these gauge choices and still obtain the same conclusions from the 
linearization stability analysis, a result which will make easier any generalization of these results.

The paper is organized as follows. In Section II we outline our perturbative approximation, and in Sections II A and B we outline our classical gauge-fixing procedure for the 
perturbations and discuss quantum anomalies associated with imposing them along with the equations of motion. We then 
briefly introduce the linearization instabilities associated with the leading order approximation of the field equations in Section III and discuss the quantum 
linearization instability constraints and their quantum anomalies with respect to the equations of motion and our gauge conditions.  Section IV ends the 
paper with a discussion and conclusion. 

\section{ Perturbative approximation }

We begin by perturbing the usual Einstein field equations 
\begin{eqnarray}
G_{ab} (g_{ab}) = \kappa T_{ab} (g_{ab}, \phi), 
\end{eqnarray}
where $\kappa \equiv 8 \pi G$, about the de Sitter solution 
\begin{eqnarray}
\nonumber
ds^2 = \bar{g}_{ab} dx^{a} dx^{b} = -dt^2 + \cosh(t)^2 (d\chi^2 + \sin(\chi)^2 d{\Omega}(\theta, \eta)^2)
\end{eqnarray}
to second order in a small parameter $\epsilon$. Here $t, \chi, \theta, \eta$ are the usual coordinates of the closed chart covering de Sitter via $\Re \times S^3$, where the 
comoving time is $t$ and the angles $(\chi, \theta, \eta)$ represent the angles of $S^3$, and $d\Omega^2$ is the $S^2$ metric $ds^2 = d \theta^2 + \sin(\theta)^2 d \eta^2$.  We 
write our perturbation ansatz as 
\begin{eqnarray}
g_{ab} &=& \bar{g}_{ab} (t, \chi, \theta, \phi) + \epsilon^2 \delta^{2} g_{ab} (t, \chi, \theta, \eta) \\
\phi &=& \bar{\phi} + \epsilon \delta \phi (t, \chi, \theta, \eta),  \ \ \bar{\eta} \in \Re, 
\end{eqnarray}
i.e. the leading order gravitational fluctuations are taken to occur at {\it second} order in $\epsilon$, with background quantities denoted by overbars, and the scalar field 
is denoted by $\phi$. In other words we 
are ignoring the transverse-traceless (gravity wave) excitations of the metric which could occur at first order in $\epsilon$.  Note that because we assume that the first order 
metric perturbations are zero, the first order coordinate transformations are restricted to Killing vector transformations of the background spacetime. With this approximation 
the leading order gravitational perturbation of the de Sitter background can be symbolically written as
\begin{eqnarray}
\ \ \ {}^{ {} }  \ \ \ \  \underbrace{ {\cal{L}}  [\delta^2 g_{ab}] }_{\mbox{ Linearized gravity } }  
			= \underbrace{ \kappa {\cal{Q}}_{ab} [ (\delta \phi)(\delta \phi) ] }_{\mbox{ Nonlinear source } },
\end{eqnarray} 
where ${\cal{L}}$ is a {\it linear} second order hyperbolic operator and $ {\cal{Q}}_{ab} $ is an operator of mixed character acting on the quadratic collection of 
matter fluctuations. We furthermore assume that the fluctuations $\delta \phi$ are {\it massless}, i.e. that they satisfy 
\begin{eqnarray}
 \bbbox \delta \phi = 0, 
\end{eqnarray}
where $\bbbox \equiv \nablauc \nabladc$. 
One can show that 
\begin{eqnarray}
\nonumber
 {\cal{L}}  [\delta^2 g_{ab}]  &=& 
( \bbbox + \frac{ 2 \Lambda }{3} ) \delta^{2} g_{ac} + ( \frac{ \Lambda \bar{g}_{ac} }{3} - \nablada \nabladc + \bbbox \bar{g}_{ac} ) \delta^{2} g
			- \bar{g}_{ac} \nablauell \nablaum \delta^{2} g_{\ell m} \\ 
&& + 2 \bar{\nabla}_{(a} \bar{ \nabla }^{m} \delta^{2} g_{c)m},   
\end{eqnarray}
where $\delta^2 g \equiv \bar{g}^{ab} \delta^{2} g_{ab}$ is the trace of the metric perturbation and where the standard summation convention applies for repeated indices.  The 
right hand side of equations (4) is simple enough to derive by inspection of equation (1). The result is
\begin{eqnarray}
\kappa {\cal{Q}}_{ab} [ (\delta \phi)(\delta \phi) ] 
			&=& 2 \kappa ( \nablada \delta \phi \nabladc \delta \phi - \frac{ \bar{g}_{ac} }{2} \nabladm \delta \phi \nablaum \delta \phi ),
\end{eqnarray}
and thus equations (6) and (7) combined in equations (4), along with the matter equation (5), form the combined second order Einstein-matter equations we wish to solve. 

As usual, the solutions $\delta^2 g_{ab}, \delta \phi$ to these equations will be invariant under a group of coordinate transformations at second order which are referred to as 
gauge conditions in cosmological perturbation theory.  We can strategically 
choose the so-called gauge condition to simplify either the interpretation of the fluctuations or the amount of work required in getting to the solutions. In the next section we
outline a gauge choice which simplifies equations (1) themselves.  It is immediately worth noting, however, that because of the vacuum de Sitter background, the scalar field 
perturbation $\delta \phi$ will be gauge-invariant to first order.

\subsection{ Gauge fixing conditions }

We show in this section that it is possible to transform to a system of coordinates wherein the divergence
of the metric perturbation, $\nablaua \delta^2 g_{ab}$, obeys a simple condition which renders the trace of the field equations (4) free of any matter terms. However, this 
choice still leaves some residual second order coordinate freedom which one can use. We show that we can use some of this freedom to furthermore make the trace of the 
metric vanish, however these particular gauge choices are in no way essential to obtain the linearization stability constraints of this paper but may be relevant for 
future generalizations of our results.

The condition on the divergence we choose is 
\begin{eqnarray}
\nablaum \delta^{2} \hat{g}_{\ell m} - \frac{ \nabladell }{2} ( \delta^{2} \hat{g} )&=& \frac{ \kappa}{2} \nabladell ( \delta \phi )^2, 
\end{eqnarray}
where the carrot denotes that these particular conditions have been fixed. The main advantage of this gauge choice is that equations (4) assume the simple form  
\begin{eqnarray}
( \nablauc \nabladc - \frac{2 \Lambda }{3} ) \delta^2 \hat{g}_{ab}  +  \frac{2 \Lambda }{3} \bar{g}_{ab} \delta^{2} \hat{g} 
												&=& -2 \kappa \delta \phi \nablada \nabladb \delta \phi, 
\end{eqnarray}
where $\delta^2 \hat{g} \equiv \bar{g}^{ab} \delta^2 \hat{g}_{ab}$.
The perturbed metric $\delta^2 \hat{g}_{ab}$ in equations (8) is related to any other perturbed metric via the Lie transformation law
$\delta^2 \hat{g}_{ab} = \delta^2 g_{ab} + \lie_{\xi_{(T)}} \bar{g}_{ab}$, where $\xi_{T}^{a}$ is an infinitesimal vector field which represents the (transverse) gauge 
transformation at second order in $\epsilon$. Here, $\lie_{\xi_{(T)}} \bar{g}_{ab}$ is the usual Lie derivative of the background metric along $\xi^{\beta}_{(T)}$. Setting the 
conditions (8) is thus equivalent to solving for $\xi^{\beta}_{(T)}$ from the equations
\begin{eqnarray}
\frac{1}{2} (\bar{\nabla}^{b} \bar{\nabla}_{b} + \Lambda) \xi_{a}^{(T)} &=& -\bar{\nabla}^{b} \delta^2 g_{ab}
									+ \frac{ \kappa }{2} \nablada (\delta \phi )^2 + \frac{ \nablada }{2} \delta^{2} g.
\end{eqnarray}
We can always find a smooth solution to equation (10), which is hyperbolic, with given smooth initial data.  We now wish to find a solution which not only satisfies
equation (10) but which also makes the {\it trace} of the resultant metric equal to zero. In other words the gauge transformation must satisfy 
\begin{eqnarray}
2 \nablada \xi_{(Tr)}^{a} = - \delta^2 \hat{g}
\end{eqnarray}
in addition to equation (10). It can be easily verified that the simple solution 
\begin{eqnarray}
\xi_{(Tr)}^{a} = \frac{1}{4 \Lambda} \nablaua \delta^2 \hat{g}
\end{eqnarray}
satisfies both requirements using the equation 
\begin{eqnarray}
( \bbbox + 2 \Lambda ) \delta^{2} \hat{g} = 0,
\end{eqnarray}
via equation (5). We note that 
\begin{eqnarray}
[ \nablada, (\bbbox + \Lambda)  ]  \xi^{a}_{(R)} =  \Lambda \nablada \xi^{a}_{(R)},
\end{eqnarray}
which implies that $( \bbbox + 2\Lambda ) \nablada \xi^{a}_{(R)} = 0$ hold iff $( \bbbox + \Lambda ) \xi^{a}_{(R)} = 0$, i.e. the residual coordinate transformations effecting 
tracelessness cannot undo transverseness and are preserved under evolution. Thus we can assume that the metric $\delta^2 \hat{g}^{\prime}_{ab}$ obeys both the transversality 
condition and tracelessness, where the prime denotes the latter property. Finally, the field equations in this gauge are 
\begin{eqnarray}
( \nablauc \nabladc - \frac{2 \Lambda }{3} ) \delta^2 \hat{g}^{\prime}_{ab} &=& -2 \kappa \delta \phi \nablada \nabladb \delta \phi.
\end{eqnarray}
We note in passing that the absence of matter terms in the trace of the field equations allows for further simplifications, and in particular we still have some gauge freedom 
remaining. If we add  $\nablada ( \xi^{(R)}_{b}) + \nabladb ( \xi^{(R)}_{b})$ to $\delta \hat{g}^{\prime}_{ab}$ such that 
\begin{eqnarray}
\nablaum \delta^{2} \hat{g}^{\prime}_{\ell m} &=& \frac{ \kappa}{2} \nabladell ( \delta \phi )^2 \\
\delta^2 \hat{g}^{\prime} &=& 0,
\end{eqnarray}
i.e. such that 
\begin{eqnarray}
(\bbbox + \Lambda ) \xi_{a}^{(R)} &=& 0, 
\end{eqnarray}
then the resultant metric will still be traceless and obey equation (8). In this paper we will not consider any further gauge-fixing on the metric since it does not bear on our 
main linearization stability result, though presumably any generalization of this work will have to confront this issue. 

\subsection{ Quantum anomalies in gauge-fixing  }

Formally, all of the above expressions containing `Wick monomials' $(\delta \phi)^2$ and $\delta \phi \nablada \nabladb \delta \phi$ are intrinsically meaningless without 
some renormalization or regulation to treat the infinities, since the operator $\delta \phi$ is a distribution. After applying a given renormalization scheme it is likely that 
not all of the gauge conditions, equations of motion, or any further conditions, can hold simultaneously {\it especially} if one furthermore demands that they be local or 
covariant in the sense of \cite{Hollands:2004yh}. For example, Hollands and Wald prove that for a massless, free quantum field $\delta \phi$ 
satisfying the linear equation of motion
\begin{eqnarray}
\bbbox \delta \phi = 0
\end{eqnarray} 
it is {\bf not} in general possible to {\it also} satisfy the nonlinear conditions
\begin{eqnarray}
\delta \phi \bbbox \delta \phi &=& 0 \\
( \nabladb \delta \phi ) \bbbox \delta \phi &=& 0
\end{eqnarray}
if one insists on $\delta \phi$ being a local and covariant quantum field. In this sense a quantum anomaly is said to occur. It is worth noting that equation (20) 
must be valid if the tracelessness condition below equation (11) holds, and that equation (21) is the requirement of stress energy conservation, so in other words Hollands 
and Wald showed that any local and covariant (massless) quantum fluctuation $\delta \phi$ would possess a quantum anomaly in this gauge fixing. Specifically, they   
established a set of conditions involving local curvature scalars which one would have to satisfy in order to avoid these sorts of anomalies, and they showed that for general 
spacetimes these curvature conditions cannot be satisfied.  However, it turns out that the Hollands and Wald (HW) anomaly requirement does not hold for perturbation theory 
about the maximally symmetric de Sitter spacetime because, e.g., all covariant derivatives of the curvature are zero. In other words, the conditions they derived which forbid 
the simultaneous satisfaction of the auxiliary conditions (20) and (21) with the equation of motion are actually satisfied for a de Sitter background spacetime.

Indeed, if we simply assume a renormalization prescription which satisfies 
locality and covariance in the sense of HW, then our Klein-Gordon stress-energy may be written exclusively in terms of the `Wick monomials` $\Psi \equiv ( \delta \phi )^2$, 
$\Psi_{ab} \equiv \delta \phi \nablada \nabladb \delta \phi$ (see HW in \cite{Hollands:2004yh}, their Section 3.2 ): 
\begin{eqnarray}
T_{ab} &=& \frac{ 1}{2} \nablada \nabladb \Psi - \Psi_{ab} - \frac{\bar{g}_{ab}}{4} \bbbox \Psi,  
\end{eqnarray}
which, using the fact that $\bar{g}^{\ell a} [ \nabladell, \nabladb] \nablada \Psi = \bar{R}_{\ell b} \nablauell \Psi$, implies that\footnote{ We are here freely using the 
Leibniz condition, i.e. asserting that $\nablada \Psi = 2 \delta \phi \nablada \delta \phi$. This condition forms part of the locality and covariance requirements of HW, so that
in a direct sense the question of whether or not anomalies exist is also a question of whether or not the Leibniz condition can be imposed along with the equation of motion 
$\bbbox \delta \phi = 0$.} 
\begin{eqnarray}
T^{a}_{a} &=& \frac{1}{2} \bbbox \Psi - \Psi^{a}_{a} - \bbbox \Psi, \\
\Psi^{a}_{a} &=& \delta \phi \bbbox \delta \phi, \\
\nablaua T_{ab} &=& \frac{1}{2} \bar{R}_{\ell b} \nablauell \Psi + \frac{1}{4} \nabladb \bbbox \Psi - \nablaua \Psi_{ab} =  \nabladb \delta \phi \bbbox \delta \phi
\end{eqnarray}
It turns out that one can calculate these quantities using a so-called Hadamard normal ordering prescription 
(see \cite{Hollands:2004yh} and reference [17] therein for details) and we simply quote the result:
\begin{eqnarray}
\delta \phi \bbbox \delta \phi &=& Q, \\
\nabladb \delta \phi \bbbox \delta \phi &=& \frac{1}{3} \nablada Q, 
\end{eqnarray}
where $Q$ is a nonvanishing local curvature scalar. The main point is that if one wants to ensure that the left hand sides of equations (25) and (24) vanish, then one must 
redefine $\Psi$ and $\Psi_{ab}$ in a manner consistent with `locality` and `covariance`. As HW have proven in \cite{Hollands:2001fb}, the freedom one has in doing this is 
actually fairly restrictive and amounts to the transformations
\begin{eqnarray}
\Psi &\rightarrow& \Psi + C \\
\Psi_{ab} &\rightarrow& \Psi_{ab} + C_{ab}, 
\end{eqnarray}
where $C$ is any scalar constructed out of the metric, curvature, and derivatives of the curvature with dimension $[length]^{-2}$ and $C_{ab}$ is any symmetric tensor that 
is similarly constructed, with dimension $[length]^{-4}$. Therefore, if we try to use the available freedom given by equations (28) and (29) to make the 
left hand sides of equations (26) and (27) vanish, we obtain the conditions, using equations (24) and (25),
\begin{eqnarray}
 \frac{1}{2} R_{\ell b} \nablauell (\Psi + C) + \frac{1}{4} \nabladb \bbbox (\Psi + C) - \nablaua (\Psi_{ab} + C_{ab})  &=& 0 \\
  (C^{a}_{a} + \Psi^{a}_{a}) &=& 0
\end{eqnarray}
which are equivalent to
\begin{eqnarray}
 \frac{1}{2} R_{\ell b} \nablauell C + \frac{1}{4} \nabladb \bbbox C - \nablaua C_{ab} &=& - \frac{1}{3} \nabladb Q \\
 C^{a}_{a} &=& -Q.
\end{eqnarray}
Since de Sitter spacetime is maximally symmetric, the most general form of $C$ can be $C = \alpha R$, $\alpha \in \Re$, since $R$ is the only natural 
quantity with dimensions $[length]^{-2}$ for a massless, minimally coupled, scalar field. Putting this form in for $C$ we obtain the equations
\begin{eqnarray}
- \nablaua C_{ab} &=& - \frac{1}{3} \nabladb Q \\
C^{a}_{a} &=& -Q, 
\end{eqnarray}
which has the obvious solutions $C_{ab} = n g_{ab} Q + \beta g_{ab} R^2$, $n, \beta \in \Re$. Putting this in, it is simple to obtain the conditions
\begin{eqnarray}
n &=& \frac{1}{3} \\
Q &=& -\frac{12}{7} \beta R^2
\end{eqnarray}
In  \cite{Hollands:2001fb} and references therein it is shown that $Q$ can only be a function of curvature invariants and their derivatives, which for maximally symmetric 
spacetimes reduces to a linear combination of $R^2$ terms for dimensional reasons. Therefore it is always possible to pick a particular real value of $\beta$ to satisfy 
equation (37). Equation (36) shows that picking  $n = \frac{1}{3}$ satisfies equation (34) as well, so that in total we can satisfy (30) and (31) 
{\it simultaneously} in the maximally symmetric de Sitter spacetime. 

The simultaneous satisfaction of equations (30) and (31) ensures that we can impose the conditions $\nabladb \delta \phi \bbbox \delta \phi = 0$ and 
$\delta \phi \bbbox \delta \phi = 0$ in addition to the equation of motion $\bbbox \delta \phi=0$ while {\it also} insisting that the quantum field $\delta \phi$ is local 
and covariant in the sense of HW. Since the former two conditions are equivalent to stress-energy conservation and the tracelessness condition of the previous section. 
Thus we have demonstrated that there exist no quantum anomalies with respect to those conditions and the equation of motion 
$\bbbox \delta \phi = 0$ using the formalism developed by HW. This is equivalent to the claim that there are no quantum anomalies in our gauge fixing. We may now ask if it is
possible to impose additional conditions, in particular so-called linearization stability conditions, in a consistent way. 

\section{ Linearization instability }

Although commonly presented in the language of the initial value constraints on a given spacelike slice of spacetime, the appearance of linearization 
instabilities can be seen in a more straightforward way which naturally emphasizes the full four dimensional equations. The Bianchi identity 
\begin{eqnarray}
( G_{a}^{b} + \Lambda \delta_{a}^{b} )_{;b} &=& 0, 
\end{eqnarray}
always holds for any metric.  If one varies this relation with respect to the metric then around the de Sitter background we have
\begin{eqnarray}
\left( \overrightarrow{ \frac{\delta G_{a}^{b}}{\delta^2 g_{\ell m}} } (\delta^2 g_{\ell m}) 
+ \overrightarrow{ \frac{\delta (\Lambda \delta^{a}_{b})}{\delta^2 g_{\ell m}}} (\delta^2 g_{\ell m}) \right)_{;b} + \delta^2 \Gamma^{b}_{\ell b} \bar{G}_{a}^{\ell} 
											- \delta^2 \Gamma^{\ell}_{b a} \bar{G}_{\ell}^{b} &=& 0, 
\end{eqnarray}
where the notation $\overrightarrow{ \frac{\delta M}{ \delta^2 g_{ab}} } (\delta^2 g_{ab})$ means the first variation of the function $M$ with respect to $\delta^2 g_{ab}$
written as a linear operator acting on $\delta^2 g_{ab}$. Similarly, 
$2 \delta^2 \Gamma^{a}_{bc} \equiv \bar{g}^{da} ( \delta^2 g_{dc;b} + \delta^2 g_{db;c} - \delta^2 g_{bc;d} )$ is the perturbed Christoffel symbol. 
Since for our case the background $\bar{G}^{a}_{b} + \Lambda \delta^{a}_{b} = 0$, then the latter terms involving the perturbed Christoffel symbols drop out and we are left with 
simply 
\begin{eqnarray}
\left( \overrightarrow{ \frac{\delta G_{a}^{b}}{\delta^2 g_{\ell m}} } (\delta^2 g_{\ell m}) \right)_{;b} &=& 0
\end{eqnarray} 
for arbitrary $\delta^2 g_{\ell m}$. Notice how the cosmological constant term has dropped out of equation (39)
Now, we can similarly vary the quantity $\left( X^{a}  G_{a}^{b}  + \Lambda \delta^{a}_{b}  \right)_{;b}$ with respect to $g_{ab}$ and obtain 
\begin{eqnarray}
\left( X^{a}  \overrightarrow{ \frac{\delta G_{a}^{b}}{\delta^2 g_{\ell m}} } (\delta^2 g_{\ell m})  ) \right)_{;b} 
+ \left(  X^{a} G_{a}^{b}  + \Lambda X^{a} \delta^{b}_{a} \right)_{\delta^2 ;b} &=& X^{a}_{;b}  
								\overrightarrow{ \frac{\delta G_{a}^{b}}{\delta^2 g_{\ell m}} } (\delta^2 g_{\ell m}),
\end{eqnarray}
which is zero if $X^{a}$ is a Killing vector. The last term of the right hand side denotes the variation of the Christoffel symbols implicit in the covariant derivative. 
Using Gauss' divergence theorem and the fact that the spatial sections have no boundary, 
$\int  X^{a} n_{b} \overrightarrow{ \frac{\delta G_{a}^{b}}{\delta^2 g_{\ell m}} } (\delta^2 g_{\ell m}) \sqrthb d^{3} x$ is independent of 
the hypersurface and independent 
of the variation $\delta^2 g_{\ell m}$ (here, $n^{a}$ is the normal to the spacelike hypersurface). Since the variation is arbitrary, we can choose 
$\delta^2 g_{ab}$ to be zero on one of the hypersurfaces, and thus, we have in general  
\begin{eqnarray}
\int  X^{a} n_{b} \overrightarrow{ \frac{\delta G_{a}^{b}}{\delta^2 g_{\ell m}} } (\delta^2 g_{\ell m}) \sqrthb d^{3} x = 0
\end{eqnarray}
holds for {\it arbitrary} variations. However, we want the second order fluctuations $\delta^2 g_{ab}$ to obey 
\begin{eqnarray}
\overrightarrow{ \frac{\delta G^{a}_{b}}{ \delta^2 g_{cd}} } (\delta^2 g_{cd} ) &=& \kappa T^{a}_{b} ( \delta \phi, \delta \phi), 
\end{eqnarray}
equation (15), where the right hand side is the stress-energy tensor to second order in $\delta \phi$. Note that the variation is of the mixed valence tensor $G^{a}_{b}$, so 
that the variation of the cosmological constant term $\Lambda \delta^{a}_{b}$ is zero. Thus we must have 
\begin{eqnarray}
\int n_{a} X^{b} T^{a}_{b} (\delta \phi, \delta \phi) \sqrthb d^3 x &=& 0, 
\end{eqnarray}
bearing in mind equation (22). This is an integral constraint on the fluctuation $\delta \phi$ imposed because of our demand that the second order equations have a solution 
for the metric fluctuations, is precisely the so-called {\it linearization instability constraint}. 

One can also carry out the above argument in the Hamiltonian formalism, since the above expressions are exactly the initial value constraint equations, and since we will 
want to express the LS conditions in terms of quantum operators.  Indeed, it is possible to characterize the total gravitational field in terms of the three-geometry 
$h_{ij}$ at some time $t$, and its conjugate momentum $\pi^{ij}$, of a compact spacelike surface $\Sigma_{t} = S^{3}$ (see Wald in \cite{wald}). 
Linearizing the conjugate pair in our notation, their commutation relations can be written as 
\begin{eqnarray}
\nonumber
[\delta^2 h_{ab}(x) , \delta^2 \pi^{cd}(\xp)] &=& i \sqrthb \left[ \delta^{c}_{a} \delta^{d}_{b} + \delta^{d}_{a} \delta^{c}_{b} \right] 
									\delta (\chi^{\prime} - \chi) \delta( \theta^{\prime} - \theta) \delta(\eta^{\prime} - \eta) \\
&\equiv& i \sqrthb \left[ \delta^{c}_{a} \delta^{d}_{b} + \delta^{d}_{a} \delta^{c}_{b} \right] \delta^{3}( x - \xp), 
\end{eqnarray}
and similarly $[\delta \phi, \delta \pi_{\phi}] = \sqrthb \delta^{3} (x - \xp)$. The classical metric and matter fluctuations are not free on $S^3$, of course, 
but satisfy the Hamiltonian and momentum constraints order by order in $\epsilon$. The {\it matter parts} of the constraints are 
\begin{equation}
\left . 
\begin{array}{rcl}
\delta^2 {\cal{H}}_{\perp} &=&  \bar{N} \frac{(\delta{\pi}_{\phi})^2}{\sqrt{|\bar{h}|}}  
				+ \bar{N} \frac{ \sqrt{|\bar{h}|} }{2}  \bar{D}_{i} \delta \phi \bar{D}^{i} \delta{\phi}  \\
\delta^2 {\cal{H}}^{i} &=& - \kappa \delta \piphi \lie_{\vec{\bar{N}}} \delta \phi \\
\delta \piphi &\equiv& -\frac{1}{2} \bar{N} \sqrthb \left[ 2 \dot{\delta \phi} \left[ -\frac{1}{\bar{N}^2} \right] 
							+ 2 \left[  \frac{\bar{N}^{i}}{\bar{N}^2} \right] \delta \phi_{,i} \right]
\end{array} \right\}
\end{equation}
Here, the $\bar{D}_{a}$ are covariant derivatives 
associated with $\bar{h}_{ab}$, $\bar{\Delta} \equiv \bar{D}^{a} \bar{D}_{a}$, and all indices on three dimensional objects are spatial only. The background lapse, $\bar{N}$, 
and shift, $\vec{\bar{N}}$, are $1$ and $\vec{0}$ respectively by our choice of slicing.

Finally, in terms of the above $3+1$ language, the LS conditions we wish to consider are 
\begin{eqnarray}
 2 \kappa \intsigma \left\{ \left(  \sqrthb \left[ \frac{1}{2}\bar{D}^{i} \bar{D}_{i} (\delta \phi)^2  - \delta \phi \bar{D}^{i} \bar{D}_{i} \delta \phi \right]   
	+ \frac{(\delta \pi_{\phi})^2}{2\sqrthb}  \right) X_{\perp}  
		     - \frac{1}{2} \delta \pi_{\phi} \lie_{\vec{X}} \delta \phi \right\} d^{3}x &=& 0,  
\end{eqnarray}
using the Leibniz rule. Here $X_{\perp} \equiv \bar{n}^{a} X_{a}$ and $\vec{X}$ are the tangential components of 
the Killing vector field $X^{a}$, where $\bar{n}^{a}$ is a unit normal to the hypersurface $\sigmat$, and we note that equation (47) is just the integral of the sum of the 
matter parts of the constraints (46) with $(\bar{N}, \bar{N}^{i})$ replaced by the Killing vector $(X_{\perp}, X^{i})$.  
It can be shown that the these LS conditions are gauge-invariant and conserved from hypersurface to hypersurface, and furthermore there are {\it ten} of them, one for each of 
the Killing isometries of the de Sitter background. 

One may further simplify the LS conditions (47) by expressing them in terms of the quantities $\Psi$ and $\Psi_{ab}$ of Section II B.
Indeed, in the context of our slicing, it is easy to see that the conjugate momentum of the scalar field is 
\begin{eqnarray}
\delta \pi_{\phi} &=& \sqrthb \bar{\nabla}_{0} \delta \phi, 
\end{eqnarray}
which when combined with an integration by-parts in our topologically closed slicing leads to 
\begin{eqnarray}
\intsigma \delta \pi_{\phi} \lie_{X} \delta \phi d^3 x &=&  - \intsigma \sqrthb \left[ (\Psi_{ia}  - \bar{n}_{a} \frac{3 H}{2} \nabladi \Psi ) X^{i}
											+ \frac{1}{2} ( \nablada \Psi ) \bar{D}_{i} X^{i} \right] \bar{n}^{a} d^3 x,
\end{eqnarray}
where we remind the reader that the definitions  
\begin{eqnarray}
\Psi_{ab} \equiv \delta \phi \nablada \nabladb \delta \phi \\
\Psi \equiv (\delta \phi)^2
\end{eqnarray}
apply to the {\it four} dimensional background connection and associated derivative operator $\nablada$. Applying the Leibniz rule to the momentum term we find 
\begin{eqnarray}
\intsigma  \frac{(\delta \pi_{\phi})^2}{2 \sqrthb} X_{\perp} d^{3} x&=& 
\intsigma \frac{\sqrthb}{2} \left[ -\Psi_{ab} + \frac{1}{2} \partial_{a} \partial_{b} \Psi \right] X_{\perp} \bar{n}^{a} \bar{n}^{b} d^3 x,
\end{eqnarray}
and finally the remaining terms are simply
\begin{small}
\begin{eqnarray}
2 \kappa \intsigma \sqrthb \left(  \frac{1}{2}\bar{D}^{i} \bar{D}_{i} (\delta \phi)^2  - \delta \phi \bar{D}^{i} \bar{D}_{i} \delta \phi \right) X_{\perp} d^3 x
&=& 2 \kappa \intsigma \sqrthb \left(  \frac{1}{2}\bar{D}^{i} \bar{D}_{i} \Psi  - \bar{h}^{ij} \Psi_{ij} - \frac{3H}{2} (\partial_{a} \Psi) \bar{n}^{a} \right) X_{\perp} d^3 x,
\end{eqnarray}
\end{small}
where the last term proportional to $H$ comes from expressing three dimensional derivative terms ($\bar{D}_{i}$) in terms of the spatial parts of four dimensional 
derivative ($\nabladi$) terms\footnote{ Thus for example 
$\delta \phi \bar{D}^{i} \bar{D}_{i} \delta \phi = \delta \phi \nablaui \nabladi \delta \phi + \delta \phi \bar{h}^{ij} \bar{\Gamma}^{0}_{ij} \nabladO \delta \phi 
= \Psi^{i}_{i} + \frac{3H}{2} \partial_{0} \Psi$}. 

Putting the above expressions into equation (47) we obtain the LS conditions exclusively in terms of projections $\Psi_{ab}$ and $\Psi$
\begin{small}
\begin{eqnarray}
\nonumber
2 \kappa && \intsigma \sqrthb \left\{  \left[ \frac{1}{2}\bar{D}^{i} \bar{D}_{i} \Psi  
- \bar{h}^{ij} \Psi_{ij} - \frac{3H}{2} (\partial_{a} \Psi) \bar{n}^{a} \right]  X_{\perp}  
	+   \frac{1}{2} \left[ -\Psi_{ab} + \frac{1}{2} \partial_{a} \partial_{b} \Psi \right] X_{\perp} \bar{n}^{a} \bar{n}^{b} \right .  \\
&&
\left . 
+  \left[  (\Psi_{ia}  - \bar{n}_{a} \frac{3 H}{2} \nabladi \Psi ) X^{i} + \frac{1}{2} ( \nablada \Psi ) \bar{D}_{i} X^{i} \right] \bar{n}^{a}      \right\} d^{3}x 
\stackrel{*}{=} 0,
\end{eqnarray}
\end{small}
where $H \equiv \partial_{0} \ln (a(t))$ and $3H^2 = \kappa \Lambda$. It is important to emphasize that this form of the LS conditions, compared to equations (47), is 
less general in the sense that one can still have a compact slicing of de Sitter without demanding that the background metric is diagonal--that is why we have labelled 
equation (54) with an overstar.

The primary utility of equation (54) is its notation, since naturally leads to the question of whether or not quantum anomalies 
exist with respect to it, the conditions (30) and (31), and the linear equation of motion for $\delta \phi$. 

\subsection{ Quantum anomalies in the LS conditions }

If anomalies exist in the imposition of the additional condition (52) then their primary mathematical effect would be to add a source term to the right hand side of (52), 
which might imply a more exotic constraint, especially if the source is large in some sense (and thus forces the fluctuations to be large).  Therefore it is 
important to check whether or not anomalies 
really do exist for our special case of de Sitter, although a-priori there are considerable grounds for optimism in the case of de Sitter since the anomalies are all constants.

Returning to the HW formalism, we once again regard the quantity $\delta \phi$ as a quantum operator and seek to redefine the quantities $\Psi$ and $\Psi_{ab}$ in a manner 
consistent with the HW axioms which define $\Psi, \Psi_{ab}$ to be local and covariant (again, see \cite{Hollands:2001fb} for a proof of uniqueness of $\Psi, \Psi_{ab}$ 
up to local curvature terms of the right dimension). This re-definition was already used in equations (28) and (29), and for the case of the quantum analogue of 
equation (54) it amounts to writing
\begin{small}
\begin{eqnarray}
\nonumber
2 \kappa  && \intsigma \sqrthb \left\{  \left[ \frac{1}{2}\bar{D}^{i} \bar{D}_{i} (\Psi + C)  
- \bar{h}^{ij} (\Psi_{ij} + C_{ij}) - \frac{3H}{2} (\partial_{a} (\Psi + C)) \bar{n}^{a} \right]  X_{\perp}  \right . \\
\nonumber
&&	
+   \frac{1}{2} \left[ -(\Psi_{ab}+C_{ab}) + \frac{1}{2} \partial_{a} \partial_{b} (\Psi + C) \right] X_{\perp} \bar{n}^{a} \bar{n}^{b} \\
&&
\left . 
+  \left[  ( (\Psi_{ia}+C_{ia})  - \bar{n}_{a} \frac{3 H}{2} \nabladi (\Psi + C) ) X^{i} 
					+ \frac{1}{2} ( \nablada (\Psi + C) ) \bar{D}_{i} X^{i} \right] \bar{n}^{a}      \right\} d^{3}x 
\stackrel{*}{=} 0,
\end{eqnarray}
\end{small}
which simplifies drastically for the case of our de Sitter background since $C$ is a constant and $C_{ab} \propto \bar{g}_{ab}$. Thus, the quantum anomaly terms immediately 
reduce to an integral of the form  
\begin{eqnarray}
2 \kappa \intsigma \sqrthb \left\{   \left[  - \bar{h}^{ij}  C_{ij} \right]  X_{\perp}  
			+   \frac{1}{2} \left[ -C_{ab} \right] X_{\perp} \bar{n}^{a} \bar{n}^{b} \right\} d^{3}x =  - 5 \kappa  Q \intsigma X_{\perp} \sqrthb d^{3}x.
\end{eqnarray}
However, it is a fact that $\int X_{\perp} \sqrthb d^3 x = 0$ for all of the Killing vectors of the background de Sitter spacetime (including those Killing vectors which 
have a nontrivial timelike component). In other words, no anomalies appear in the LS conditions because they are expressed as an integral over $S^3$ of the constant anomaly 
terms $C, C_{ab}$, multiplied by $X_{\perp}$, which is an odd function of space. 

We conclude that the LS 
conditions (56) do {\it not} exhibit any quantum anomalies with respect to the given coordinate conditions, the equations of motion, and the requirements of locality and 
covariance in the sense of HW. They do form a nontrivial operator constraint on the quantum states $|\Psi>$ which the operators $\delta \phi$ and 
$\delta \pi_{\phi}$ act on. We emphasize that this is the same conclusion that Higuchi reached using completely different methods in \cite{Higuchi:1991tm}, however his 
calculation involved only vacuum gravitational wave (TT) fluctuations and did not consider regularization issues or quantum anomalies as such. However, it is clear 
that he anticipated the result that even if matter fields were coupled to gravity in a de Sitter background, such as in the scenario presented in this paper, one would also 
obtain a constraint in the quantum case. 

In the next section we show that the additional condition (54) imposes de Sitter invariance on the state(s) $|\Psi>$.

\subsection{ Quantum LS conditions and de Sitter invariance of $|\Psi>$} 

Higuchi in \cite{Higuchi:1991tm} has already shown that the quadratic LS conditions for gravitational fluctuations in vacuum de Sitter demand that all the physical 
(i.e., gauge-invariant) states in linearized gravity be de Sitter invariant. He did so via an analysis that showed how the de Sitter group transformations of the classical 
mode functions related with those of the creation and annihilation operators in the quantum theory.  In this way he showed that the operators $\delta^2 P(X)$ generated 
de Sitter transformations. 

However, it turns out that one may also prove this de Sitter invariance condition for our particular backreaction problem in a more compact manner. One may appeal to the 
facts proven some time ago by Moncrief in \cite{Moncrief:1976I}, namely that the LS conditions are gauge-invariant and conserved from hypersurface to hypersurface. The proof 
of these statements is technically cumbersome but straightforward. The precise statement of these properties is 
\begin{eqnarray}
\bar{n}^{a} \nablada \delta^{2} P(X) &=& 0, \\
 \delta^{2} P(X) - \left( \tilde{  \delta^{2} P(X) } \right) &=& 0, 
\end{eqnarray}
where $\left( \tilde{  \delta^{2} P(X) } \right)$ are the LS conditions with their canonical variables transformed along some vector $\zeta^{a}$, i.e. 
$\delta^2 \pi^{ij} \rightarrow \delta^2 \pi^{ij} + \lie_{\zeta} \bar{\pi}^{ij}$, etc. In other words, the LS conditions are gauge-invariant and preserved from slice to slice. 
Similarly, one may show in a straightforward calculation that the Poisson bracket of two LS conditions satisfies
\begin{eqnarray}
\{ \delta^2 P(X_{a}), \delta^2 P(X_{b}) \} &=& A^{c}_{ab} \delta^2 P (X_{c}), 
\end{eqnarray}
where the $A^{c}_{ab}$ are some trivial structure functions. However, the fact that the Poisson bracket of two LS 
conditions returns a third to within some structure constants, 
combined with the fact that the LS conditions are separately conserved (and gauge-invariant), means that they are the so-called Hamiltonian generators of the associated 
symmetry transformations (see Moncrief in \cite{Moncrief:1978te} and also Taub in reference therein) $X^{a}$. In other words, their Poisson bracket algebra must be isomorphic 
to that of the symmetry group represented by $X^{a}$, which we have taken to be Killing vectors of de Sitter spacetime. Therefore, the constants $A^{c}_{ab}$ are related to 
the Lie algebra of the Killing fields via
\begin{eqnarray}
[X_{a}, X_{b}] &=& A^{c}_{ab} X_{c}, 
\end{eqnarray}
and the constraints $ \delta^2 P(X_{a})$ must be the generators of de Sitter transformations. This is easy to verify in practice because one may easily derive the Killing 
vectors of de Sitter spacetime and therefore find the constants $A^{c}_{ab}$. 

Returning to the quantum LS conditions, the equivalent condition to (54) which one must demand is 
\begin{eqnarray}
[ \delta^2 P(X_{a}), \delta^2 P(X_{a}) ] &=& i A^{c}_{ab} \delta^2 P(X_{c}), 
\end{eqnarray}
and given that we have proven that there are no quantum anomalies which arise in the LS conditions it is possible to find a `normal ordering` such that this relationship is 
true. This can be seen in a straightforward way by taking the above commutator using equation (54), for which only terms proportional to $X_{\perp}$ appear (i.e. no 
cross terms between spatial and termporal parts of $X^{a}$ appear) with coefficient which is the straightforward square of the left hand side of equation (56). Thus, finally, 
we may identify the LS conditions $\delta^2 P(X) |\Psi> = 0$ as equivalent to the demand that the physical states $|\Psi>$ {\it must be invariant 
under the de Sitter group SO(4,1)}. 

\section{ Conclusions }

An immediate conclusion one can draw from the de Sitter invariance requirement
\begin{eqnarray}
\delta^2 P(X) \ket{\Psi} &=& 0
\end{eqnarray}
is that it applies to the {\it entire spectrum} of states $\ket{\Psi}$. In particular, it does not just apply to the vacuum state. This should be contrasted with the limit 
in 
which the cosmological constant $\Lambda$ goes to zero, where one would expect to recover physics in Minkowski spacetime. In the case of flat spacetime, only the vacuum state 
is invariant under the Poincare group whereas any excited states break this symmetry. It would seem therefore that any dynamics in de Sitter spacetime are highly restricted 
by this requirement of de Sitter invariant states, and furthermore Higuchi in \cite{Higuchi:1991tm} shows that for the case of vacuum gravitational (TT) fluctuations the only
normalizable de Sitter invariant state is the vaccum. One cannot do much with just one allowed state, the vacuum, and it is even harder to see how the flat spacetime limit 
occurs in this context.  Moncrief, 
Higuchi, and others have described this dearth of dynamics as the `apparent rigidity' of de Sitter spacetime, which may be thought of as the 
crude `remnants` of general diffeomorphism invariance of an underlying theory of quantum gravity. 

It is also worth noting that this result also applies to a massive scalar field. In that case, there is a potential term $m^2 \Psi$ which appears in the timelike LS condition, 
potentially introducing another anomaly. However these ambiguities are constants ($\Psi \rightarrow \Psi + C$) and since the timelike LS condition is an integral over 
$X_{\perp}$, which is antisymmetric over $S^3$ as we noted already, then there can be no further anomalies and the same logic carries forward to the same final result. 

It is useful to be somewhat more specific about the vacuum in de Sitter spacetime, which is considerably richer than in flat spacetime even for a massive, minimally coupled 
scalar field. In particular, for this case, there is an infinite family of vacua which are de Sitter invariant which are usually parametrized by one complex parameter $\alpha$ 
(not to be confused with the $\alpha$ of this paper), and which can have interesting short-distance (UV) and long-distance (IR) behaviour. However, only one unique state 
is thought to reduce to the Poincare invariant vacuum of 
Minkowski spacetime in the limit of $\Lambda \rightarrow 0$ and in particular for the two-point function in the vacuum to be of the `Hadamard' form (i.e. to have the same 
singularity structure as in Minkowski). This is known as the Bunch-Davies (or `Euclidean') vacuum.  

For our massless, minimally coupled case, however, the situation is somewhat more peculiar. For example, in the inflationary context it can be shown that 
the mean squared fluctuations of $\delta \phi$ grow linearly in time with inflation, i.e. 
\begin{eqnarray}
\bra{0} \delta \phi^2 \ket{0} &\approx& \frac{H^3}{4 \pi^2} t
\end{eqnarray}
However, this expression manifestly violates de Sitter invariance (since it is not a function of the geodesic distance defined in the well-known five dimensional embedding of 
de Sitter spacetime) and is instead a function 
of comoving time alone. Indeed, if one insisted on a de Sitter invariant two-point function then one would quickly find that it is infrared divergent in the massless, minimally 
coupled limit. One can easily see why this is by recalling the second order matter Lagrangian action for our case, namely
\begin{eqnarray}
{}^{(2)}S_{M} &=& -\frac{1}{2} \int \sqrtgb \left[ \bar{g}^{ab} \delta \phi_{,a} \delta \phi_{,b} \right] d^{4}x,
\end{eqnarray}
and observing that it is invariant under the transformation $\delta \phi \rightarrow \delta \phi + \mbox{constant}$. This is simply a zero mode, and the two point function is 
ill defined because all values of the spatially constant part of $\delta \phi$ are equally probable in a de Sitter invariant state, just as for an eigenstate of the momentum 
in the quantum mechanics of a free particle. In fact this observation underlies the claim that Allen proves in \cite{Allen:1985ux}, which is that there exists no de Sitter 
invariant vacuum state for the massless, minimally coupled field.  For this reason it is often assumed that the symmetry group of the vacuum is smaller, for example that of the
O(4) subgroup of SO(4,1) which are the spatial rotations on $S^3$. In that case the two-point function has no infrared divergences 
anymore, and it is this O(4) invariant vacuum which is used in calculating (63).  In other words, one may sacrifice invariance with respect to boosts in order 
to obtain a reasonable expression for the two-point function. 

However, if one demands that the gravitational field fluctuations obey the Einstein equations with second order, quantum, scalar field fluctuations, this 
leads to a relation like (54) which imposes a further invariance requirement on all the physical states of the metric and 
matter fluctuations. If the linear quantum field theory for the scalar field is not de Sitter invariant, then the second order equations for the gravitational fluctuations 
are inconsistent, and have no solutions. This is an extremely strong condition and it is not clear whether any nontrivial physics remains for a single scalar field. Whether 
or not one can have interesting dynamics with more than one scalar field (or one scalar field plus gravitational waves) remains an open question. 

There is also the issue of how to interpret the limit as the background spacetime approaches de Sitter. At precisely de Sitter spacetime one has constraints such as (54)
while even for background spacetimes `very close' to de Sitter, such as a homogeneous background with a small scalar field velocity, there is no analagous constraint. 
Indeed, let us look at the solutions to the equation $ax - by^2 = 0$. For any $a$ not equal to $0$, there is a solution for $x$ for any value of $y$, but if $a=0$ there is a 
constraint on $y$, namely that $y = 0$. For very small $a$ the solution for $x$ with non-zero $y$ becomes quite large. Similarly, in our case, the constraint is 
$\int  X^{a} n_{b} \overrightarrow{ \frac{\delta G_{a}^{b}}{\delta^2 g_{\ell m}} } (\delta^2 g_{\ell m}) \sqrthb d^{3} x = 
\int n_{a} X^{b} T^{a}_{b} (\delta \phi, \delta \phi) \sqrthb d^3 x$. If $X^{a}$ is a Killing vector, then this left hand side is identically zero. This corresponds to the 
coefficient $a$ in the above example, and results in a constraint on the matter. If $X^{a}$ is `almost` a Killing vector 
(e.g., because the background spacetime is almost de Sitter), the solution to the constraint equation will require a very large value for some component of the second order
gravitational perturbation $\delta^2 g_{ab}$. As we have shown in a previous publication, \cite{Losic:2005vg}, such divergences in the second order gravitational perturbations 
are typical of backgrounds near de Sitter spacetime. 

Finally, the actual issue of constructing these de Sitter invariant states, and the problem of dynamics, will be treated in a 
forthcoming publication, but we also refer to Woodard in \cite{Woodard:2004ut} for extensive and much more in-depth commentary on the whole issue of whether or not one really 
wants, or can even attain, de Sitter invariance in related considerations.

\section{ Acknowledgements }

We would like to thank Vincent Moncrief for pointing out the simple derivation of the linear constraint condition in equation (44). W.G.U. would also like to thank the 
Canadian Institute for Advanced Research and NSERC during this work. B.L. thanks Emil Mottola, Don Page, and Richard Woodard for many useful discussions and 
NSERC for financial support.

\end{document}